\documentclass[12pt]{article}

\newcommand{\bbZ}{{\mathbb Z}}

 \newcommand{\zSn}{\sum _{n\neq  0} }
\usepackage{color}
\usepackage{epsfig,cmmib57}
\usepackage{amssymb,amsmath,exscale}
 \numberwithin{equation}{section}

\newcommand{\n}{\lambda_n}

\newcommand{\ZOMq}{\Omega}

\newcommand{\zben}{\beta_n}
\newcommand{\zg}{\gamma}

\newcommand{\intt}{\int_0^t}
\newcommand{\ints}{\int_0^s}
\newcommand{\intr}{\int_0^r}

\newtheorem{Theorem}{Theorem}
\newtheorem{Corollary}[Theorem]{Corollary}
\newtheorem{Lemma}[Theorem]{Lemma}

\newtheorem{Remark}[Theorem]{Remark}

\newcommand{\zdiaform}{\mbox{~~\zdia}}
\newcommand{\ZSUB}{\subseteq}
\newcommand{\ZOM}{\omega}
\newcommand{\zaa}{\alpha}
\newcommand{\ZR}{\rangle}
\newcommand{\ZL}{\langle}
\newcommand{\ZDE}{\delta}
\newcommand{\zt}{\tau}
\newcommand{\zdia}{~~\rule{1mm}{2mm}\par\medskip}
\newcommand{\zthe}{\theta}

\newcommand{\ZLA}{\label}
\newcommand{\ZIN}{\infty}
\newcommand{\zProof}{{\noindent\bf\underbar{Proof}.}\ }

\newcommand{\zzr}{{\rm I\hskip-2.1pt R}}
\newcommand{\ZBI}{\bibitem}
\newcommand{\ZD}{\;\mbox{\rm d}}
\newcommand{\zl}{\lambda}

\author{
L. Pandolfi\thanks{Dipartimento di Scienze Matematiche ``Giuseppe Luigi Lagrange'', Politecnico di Torino, Corso Duca degli Abruzzi 24, 10129 Torino, Italy (luciano.pandolfi@polito.it)}
}
\title{Sharp control time for viscoelastic bodys\thanks{
This papers fits into the research programme of the GNAMPA-INDAM and has been written in the framework of the   ``Groupement de Recherche en Contr\^ole des EDP entre la France et l'Italie (CONEDP-CNRS)''.}}

\begin{document}

\maketitle
{\bf\underline{Abstract}:} It is now well understood that equations of viscoelasticity can be seen as perturbation of wave type equations. This observation can be exploited in several different ways and it turns out that it is a usefull tool when studying   controllability.  Here we compare a viscoelastic system which fills a surface of a solid region (the string case has already been studied) with its memoryless counterpart (which is a generalized telegraph  equation) in order to prove exact controllability of the viscoelastic body at precisely the same times at which the telegraph equation is controllable.

The comparison is done using a moment method approach to controllability and we prove, using the perturbations theorems of Paley-Wiener and Bari, that a new sequence derived from the viscoelastic system is a Riesz sequence, a fact that  implies  controllability of the viscoelastic system.

The results so obtained generalize existing controllability results and furthermore show that the ``sharp'' control time for the telegraph equation and the    viscoelastic system coincide.

 \medskip


\section{Introduction}

We consider a control problem for the following equation:

 \begin{eqnarray}
\nonumber&&w_{tt} =2c w_t+\nabla \cdot(a(x)\nabla w)+q(x) w \\
\ZLA{eq:SisteMEMsecondordine} &&+\intt M(t-s)\left ( \nabla \cdot(a(x)\nabla w(s))+q(x) w(s)  \right )\ZD s +F(x,t)\,.
\end{eqnarray}  
Here $ t>0 $ and $ w\in\zzr $ denotes a function $ w(x,t) $,  $ x\in \ZOMq $  where $ \ZOMq $ is the region occupied by the body. We assume that it is a bounded region with a smooth ($ C^2 $) boundary
and $ {\rm dim}\,\Omega=d\leq 3 $. In this case, 
the physical meaning of the problem can be explained as follows: when $ \Omega\subseteq\zzr^d $ with $ d=1 $ or $ d=2 $, then $ w $ 
represents the vertical displacement of the point in position $ x $ at time $ t $ of a viscoelastic string or membrane in the linear approximation. If $ d=3 $ then the relations of Eq.~(\ref{eq:SisteMEMsecondordine}) with viscoelasticity are less stringent, since in this case the displacement   is a three dimensional vector  whose components in general are not independent so that  Eq.~(\ref{eq:SisteMEMsecondordine}) represents one of the components of the deformation only in quite special cases.

Eq.~(\ref{eq:SisteMEMsecondordine}) has to be supplemented with the initial  condition
\[ 
w(\cdot ,0)=w_0\,,\quad w_t(\cdot,0)=w_1\,.
 \]
The control $ f $ acts in the Dirichlet boundary condition:
\[ 
w(x,t)=f(x,t)\quad x\in \Gamma\ZSUB \partial\ZOMq\,,\qquad w(x,t)=0\quad x\in \partial\ZOMq\setminus \Gamma\,.
 \]
 
 We stress the fact that the control $ f $ is \emph{real valued} and we assume that it belongs to $ L^2_{{\rm loc}}(0,+\infty;L^2(\Omega)) $.

 Note that the arguments of $ w=w(x,t) $ are not explicitly indicated unless needed for clarity. According to the convenience, we shall write $ w(x,t) $ or $ w(t) $ or simply $ w $. Furthermore, $ w $ does depend on $ f $ but also this dependence is not indicated.

It is well known that (when the boundary control $ f $ is square integrable) a natural space for the evolution of the system is
\[ 
L^2(\ZOMq)\times H^{-1}(\ZOMq)
 \]
 as in the corresponding memoryless case (i.e. $ M(t)\equiv 0 $).   We choose the initial data in this space.
 
 The affine term $ F(x,t) $ depends on the history of the body for $ t<0 $. 
 
We  will study controllability of system~(\ref{eq:SisteMEMsecondordine}), i.e. we study whether, starting from every initial condition, is it possible to force $ (w(\cdot,t),w_t(\cdot,t) )$ to hit any prescribed target $ (\xi,\eta)\in L^2(\ZOMq)\times H^{-1}(\ZOMq) $ at some time $ T>0 $. 
  It is easily seen that this property does not depend on the initial condition or on the affine term. So, when studying controllability, we can assume
 \begin{equation}
 \ZLA{eq:iniCONDIperCONTRO}
w(x,0)=0\,,\qquad w_t(x,0)=0\,,\qquad F(x,t)=0 
\end{equation}
and  the problem we are studying is as follows: to investigate whether there exists a ``control time''  $ T $ such that for every $ \xi\in L^2(\ZOMq) $ and $ \eta\in H^{-1}(\ZOMq) $ there exists $ f\in L^2(0,T;L^2(\Gamma)) $ such that
  \[ 
  w(\cdot,T)=\xi\in L^2(\Omega) \,,\qquad w_t(\cdot,T)=\eta\in H^{-1}(\Omega) \,.
   \]
   
  This problem has already been studied with the methods in this paper when $ d=1 $.
  
  Note that    if controllability holds at time $ T $, it holds also at larger times. So,
  we would like to study also the infimum of the   control times (the ``sharp control time''). This is done in Section~\ref{section:SHARP} where it is proves that it coincides with the sharp control time of the (generalized) \emph{telegraph equation.} This is the special case   obtained when
   $ M(t)\equiv 0 $:
   \begin{equation}
   \ZLA{eq:telegrafisti}
   \left\{\begin{array}{l}
w_{tt} =2c w_t+\nabla \cdot(a(x)\nabla w)+q(x) w\,,\\
w(x,t)=f(x,t)\quad x\in \Gamma \,,\qquad w(x,t)=0\quad x\in \partial\ZOMq\setminus \Gamma\,.  
\end{array}\right.
\end{equation}
 Under the assumptions above on the region  $ \ZOMq $ and standard assumptions on $ \Gamma $ which we don't need to recall here, controllability holds for the telegraph equation with $ L^2 $-boundary control, as proved in~\cite{Soriano} (the definition of controllability is the same as above. See also~\cite{Shub} for controllability under distributed controls).
 
A $ \Gamma $-dependent value for the control time $ T $ of Eq.~(\ref{eq:telegrafisti})  is identified in~\cite{Soriano}.
So,   we shall study controllability of the viscoelastic body when the subset $ \Gamma\in\partial \Omega $ is so chosen that the telegraph equation is controllable at a certain time $ T $, using \emph{real} controls $ f\in
 L^2(0,T];L^2(\Gamma)) $.   
 The reason of the stress posed on the (obvious) property that the control is real will be seen below.
 
 The result we are going to prove is the following theorem:
 
 \begin{Theorem}\ZLA{teo:controllPRINCIP}
We assume:
\begin{itemize}
\item 
 $ d\leq 3 $ and $ \ZOMq\in\zzr^d  $  is a bounded region with $ C^2 $ boundary. 
\item $M(t)\in H^2 _{\rm loc}(0,+\ZIN)$  
\item $q(x)\in C(\bar \ZOMq)$ and $ a(x)\in C^1(\bar \ZOMq) $, with $ a(x)>a_0>0 $ for every $ x\in \bar\ZOMq $.
\end{itemize}

  If the telegraph equation~(\ref{eq:telegrafisti}) is controllable in time $ T $,   then   Eq.~(\ref{eq:SisteMEMsecondordine}) is controllable  at the same time $ T $. Furthermore, Eq.~(\ref{eq:SisteMEMsecondordine}) and~(\ref{eq:telegrafisti}) have the same sharp control time.
\end{Theorem}

Among the different ways in which controllability can be proved, possibly the oldest one is the reduction of a control problem to a moment problem. Theorem~\ref{teo:controllPRINCIP} has been proved using this idea when $ d=1 $ (see references below) and we are going to prove that moment methods can be used in general.

Note that once the controllability property in Theorem~\ref{teo:controllPRINCIP} has been proved, the technique in~\cite{PandDCDS2} can be applied to our system, for the identification of external distributed sources.

\subparagraph{Notations.}
We shall use the following notations. The term ``locally bounded'' denotes a sequence of (usually continuous) functions which is bounded on every compact interval (of $ [0,+\infty) $). We shall use $ \{M_n\} $ and $ \{M_n(t)\} $ to denote respectively a \emph{bounded} sequence of positive numbers and a \emph{locally bounded} sequence of positive  functions (not the same at every occurrence).

Let $ \{a_n\} $ and $ \{b_n\} $ be two sequences (in any normed space). the notation $ \{a_n\}\asymp \{b_n\} $ indicates the existence of $ m_0 >0$ and $ m_1 $ such that for every $ n $ we have
\[ 
m_0|a_n|\leq |b_n|\leq m_1|a_n|\,.
 \] 
 The numbers $ m_0 >0$ and $ m_1 $ depend on the sequences.
 
 We shall use $ \star $ to denote convolution,
 \[ 
 (f\star g)=\intt f(t-s)g(s)\ZD s\,. 
  \]

 We mentioned already that we don't indicate the space or time variables in $ w $ (or related functions, as $ \xi $, $ \eta $ and $ f $) unless needed for clarity. Furthermore, for every $ t>0 $ we introduce the notations
\[ 
Q_t=\ZOMq\times (0,t)\,,\quad G_t=\Gamma\times (0,t)\,,\quad \Sigma_t=
\left (\partial \ZOMq\right )\times (0,t)\,.
 \]
Finally,
  $ \partial/\partial\nu $ denotes normal derivative and
 \[ 
 \gamma_a\phi=  a(x)\frac{\partial \phi}{\partial \nu}  \quad \mbox{on $ \partial\Omega$ (in particular  on $ \Gamma $).}
  \]  
\subsection{References and known results}

It seems that the first nontrivial results on controllability of viscoelastic systems are due to Lasiecka and Leugering (see for example~\cite{LASIE,Leugering1984,Leugering1987}) then followed by several contribution. Among them, we consider in particular the results
 in~\cite{ZANGPRIMOLAVORO,Kim1993,PandAMO}. The paper~\cite{Kim1993} proves Theorem~\ref{teo:controllPRINCIP} (even for a nonconvolution kernel) in the case $ q(x)=0 $ and $ a(x)=1 $. More important, it explicitly assumes that the control acts on the whole boundary of $ \ZOMq $, $ \Gamma=\partial\ZOMq $ (used in the proof of~Lemma~4.1).
 
 Under these conditions the paper~\cite{Kim1993}, relaying on multiplier techniques, proves controllability when $ T\geq T_0 $, where $ T_0 $ is explicitly identified. 
 
 A similar controllability result is proved in~\cite{ZANGPRIMOLAVORO} where the proof is based on Carleman estimates. In this paper the control is distributed in a subregion close to $ \partial\ZOMq $. 
 An idea in~\cite{BarbuIannelli} can then be used to derive boundary controllability   in a smaller region,  using   controls   which acts on the whole boundary.   
 Note that in this paper the memory   kernel can be space dependent. We believe that the techniques we are going to use cannot be extended to space-dependent kernels.
  
 The paper~\cite{PandAMO} uses a different idea. The equation with memory is  considered as  a perturbation of a wave equation and its solutions are then represented using cosine operator theory (this idea is  implicit   in previous papers, for example by Leugering). It is proved in~\cite{PandAMO}   that controllability holds for the equation with memory provided that the control acts in   a part of the boundary chosen so to have controllability of the corresponding wave equation.
 However, the argument is based on category arguments and the control time is not identified.
 
 We mention that the paper~\cite{ZANGPRIMOLAVORO,PandAMO} are concerned with the heat equation with memory, i.e. with an equation of first order in time, so that they study only the controllability of the component $ w(t) $, not of the velocity, but at least the arguments in~\cite{PandAMO} are easily extended to the pair (deformation/velocity).    
 
 In conclusion, Theorem~\ref{teo:controllPRINCIP} extends and completes the results in~\cite{ZANGPRIMOLAVORO,Kim1993,PandAMO} and furthermore it uses completely different techniques, which have their independent interest: the proof uses moment methods and extends to higher space dimension the techniques and results developed 
 in~\cite{AvdBELInski,AvdoninPANDOLFI1,AvdoninPANDOLFI2,LoretiPANDOLFIsforza,PandIEOT,PandDCDS1,PandDCDS2,PandViscoUnderTRACT}.
 
 \subsection{Preliminary information}
 We report the following result (see for example~\cite{Kim2}):
  
 \begin{Theorem}
 \ZLA{teo:regoSOLUZ}
 for every $ f\in L^2(G_T)= L^2(0,T;L^2(\Gamma)) $ and every initial condition $ w(\cdot,0)=\xi \in L^2(\ZOMq) $, $ w_t(\cdot,0)=\eta \in H^{-1}(\ZOMq) $, Eq.~(\ref{eq:SisteMEMsecondordine}) admits a unique solution $ w(\cdot,t) \in C(0,T;L^2(\ZOMq))$ with $ w_t(\cdot;t)\in C(0,T;H^{-1}(\ZOMq)) $. The transformation
 \[ 
 (\xi,\eta,f)\ \mapsto\ (w,w_t)
  \]
 is linear and continuous in the indicated spaces.
 
 \end{Theorem}
 
 Let   $ A $ be the operator  in $ L^2(\ZOMq) $,
 \begin{equation}
\ZLA{eq:DefiOPeratore-A}
{\rm dom}\, A= H^2(\ZOMq)\cap H^1_0(\ZOMq)\,,\qquad Aw=\nabla\cdot\left (a(x)\nabla w\right )+q(x) w\,.
\end{equation}
This operator is selfadjoint with compact resolvent. So, it has a   normalized sequence $ \{\phi_n(x)\} $ of eigenvectors, which is an orthonormal basis of $ L^2(\ZOMq) $. We define:
\[ 
\mbox{$-\zl_n^2$ is the eigenvalue of $ \phi_n $.}
 \]
 Note the sign and the exponent, but this does not imply that $ -\zl_n^2 $ is real negative. This property does depend on the sign of $ q(x) $. But,
 $-\zl_n^2$ is real negative for  large $ n $. Moreover, the eigenvalues might not be distinct. The multiplicity of each eigenvalue is finite. A known fact is that
 \[ 
 -\zl_n^2\asymp n^{2/d}\qquad \mbox{where $ d={\rm dim}\,\ZOMq $.}
  \]
  I.e., there exists $ m_0>0 $ and $ m_1>0 $ and $ N $ such that
  \[ 
n>N\ \implies\   m_0 n^{2/d}<\zl_n^2<m_1 n^{2/d}\,,
   \]
   see~\cite[p.~192]{mikailovLIBRO}.
  So,
 \begin{Lemma}\ZLA{eq:LemmaSUconveSERIE}
 If $ d\leq 3 $ then we have $ \sum 1/\zl_n^4<+\ZIN $.
 \end{Lemma}
 This Lemma will be explicitly used in our proof.

  
  \section{\ZLA{Sect:RieszMOMENTI}Riesz sequences and moment methods}
  
  Linear control systems can be reduced to the solution of  suitable \emph{moment problems:} given a sequence $ \{ e_n\} $ in a Hilbert space $ H $ (inner product is $ \ZL\cdot,\cdot \ZR $ and the norm is $ |\cdot| $) and a sequence $ \{c_n\} \in l^2$, it is required to solve
  \begin{equation}\ZLA{eq:MOMentAstaRatto} 
  \ZL  f,e_n\ZR =c_n
   \end{equation}  
   for every index $ n $. 
   
   In general, $ \{ \ZL  f,e_n\ZR \}\notin l^2$
 and so we have an operator $ \mathbb{J} $  on $ H $, whose domain
 is
 \[ 
 {\rm dom }\,\mathbb{J}=\{ f\in H\,:\ \{ \ZL  f,e_n\ZR \}\in l^2\}\,,\quad \mbox{so that $\mathbb{J}:\ H\mapsto l^2 $.   } 
  \]
  
 In our case however, it will be $  {\rm dom }\,\mathbb{J}=H $ and $ \mathbb{J}\in \mathcal{L}(H,l^2) $.The adjoint  $ \textbf{J}^* $: $ l^2\mapsto H $ is given by
 \begin{equation}\ZLA{eq:defiJstar}
 \textbf{J}^* \left (\{v_n\}\right )=\sum e_n v_n
  \end{equation}
 
 The moment problem is to understand whether $ \mathbb{J} $ is surjective.
 
 It turns out that, when $ \mathbb{J}\in \mathcal{L}(H,l^2) $,  $ \mathbb{J} $ is surjective if and only if $ \{e_n\} $ is a \emph{Riesz sequence,} see~\cite[Theorem~I.2.1]{AvdoninIVANOV}, which is defined as follows: we first define a Riesz basis. A {\em Riesz basis\/}  of $ H $ is a complete sequence which is the image of an orthonormal basis under a linear, bounded and boundedly invertible transformation in $ H $.
 
 A sequence which is a Riesz basis in its closed span is a {\em Riesz sequence.\/}
 
 The following holds (see~\cite[Th.~9]{Young}): 
 \begin{Lemma}
The sequence $ \{e_n\} $ is a Riesz sequence if and only if there exist numbers $ m_0>0 $ and $ m_1>0 $ such that
\begin{equation}\ZLA{eq:laDEFIrieszCONDbecSUFF}
m_0\sum |a_n|^2\leq \left |\sum a_n e_n\right |^2_{H}\leq m_1\sum |a_n|^2
 \end{equation}
 for every finite sequence $ \{a_n\} $. If furthermore the sequence $ \{e_n\} $ is complete, then it is a Riesz basis.
\end{Lemma}

Every Riesz sequence admits biorthogonal sequences $ \{\psi_n\} $ i.e. sequences such that
\[ 
\ZL  \psi_k,e_n\ZR =\delta_{n,k}=\left\{\begin{array}
{lll} 
1 &{\rm if}& n=k\\
0 & {\rm if}&n\neq k\,.
\end{array}
\right.
 \]
One (and only one) of these biorthogonal sequence belongs to the closed space spanned by   $ \{e_n\} $. This biorthogonal sequence is a Riesz sequence too, and the solution of the moment problem~(\ref{eq:MOMentAstaRatto}) is 
\[ 
f=\sum c_n\psi_n\,.
 \]

 Let $ \{e_n\} $ and $ \{z_n\} $ be two sequences in $ H $. We say that they are \emph{quadratically close} if
 \[ 
 \sum |e_n-z_n|^2<+\ZIN
  \]
  and we use the following test (see~\cite{Young}):
  \begin{Theorem}\ZLA{teo:PaleyWIENERbari}
Let $ \{e_n\} $ be a Riesz sequence in $ H $ and let $ \{z_n\} $ be quadratically close to $ \{e_n\} $. Then we have
\begin{itemize}
\item {\bf Paley-Wiener Theorem:} there exists $ N $ such that $ \{z_n\} _{n>N} $ is a Riesz sequence in $ H $;
\item {\bf Bari Theorem:} the sequence $ \{z_n\} $ is a Riesz sequence if, furthermore, it is {\em $ \ZOM $-independent,\/} i.e. if (here $ \{\zaa_n\} $ is a sequence of numbers)
\[ 
\sum \zaa_n z_n=0\ \implies \{\zaa_n\}=0\,.
 \]
\end{itemize}

\end{Theorem}
 
A usefull observation is as follows: if $ \{z_n\} $ is quadratically close to a Riesz sequence then  $ \sum \zaa_nz_n $ converges in $ H $ if and only if $ \{\zaa_n\}\in l^2 $.

The concrete case we are interested in, is the case $ H=L^2(0,T;K) $ where $ K $ is a second Hilbert space (it will be $ K=L^2(\Gamma) $). 
In this context, we need two special results. 
 
\begin{Theorem}
Let $  \bbZ'=\bbZ\setminus\{0\} $ and let $ \{\zben\} _{n\in\bbZ'}$, $ \{k_n\} _{n\in\bbZ'}$ be such that 
\begin{equation}\ZLA{eq:parita}
\beta_{-n}=-\zben\,,\qquad k_n=k_{-n}\in K\,, \qquad |\mathcal{I}m\, \zben|<L \,.
 \end{equation}
 for e suitable number $ L $. If the sequence $ \{ e^{i\zben t}k_n\}_{n\in\bbZ'} $ is a Riesz sequence in $ L^2( -T,T,K) $, then the sequences
 \begin{equation}\ZLA{eq:succeSINcosDAfareRiesz}
 \left \{k_n\cos\zben t\right \}_{n>0}\,,\qquad \left \{k_n\sin\zben t\right \}_{n>0}
  \end{equation}
  are Riesz sequences in $ L^2(0,T;K) $.
   
\end{Theorem}
\zProof The assumption is that~(\ref{eq:laDEFIrieszCONDbecSUFF}) holds for the sequence $ \{ e^{i\zben t}k_n\}_{n\in\bbZ'} $ in $ L^2(-T,T;K) $. We prove that a similar property holds for the sequences~(\ref{eq:succeSINcosDAfareRiesz}) in $ L^2(0,T;K) $. We consider the cosine sequence. The sine sequence is treated analogously.

Let $ \{a_n\} $ be any finite sequence of complex numbers. Then we have
\begin{align*}
&  \left |\left |\sum_{n>0} a_n k_n\cos\zben t\right |\right |^2 _{L^2(0,T;K)}=\frac{1}{4}\left |\left | 
\sum_{n>0} a_n k_n e^{i\zben t}+\sum_{n>0} a_n k_n e^{-i\zben t}
 \right |\right |^2_{L^2(0,T;K)}\\
& =\frac{1}{8}\left |\left |  
\sum_{n>0} a_n k_n e^{i\zben t}+\sum_{n>0} a_n k_n e^{-i\zben t}
 \right |\right |^2_{L^2(-T,T;K)} 
 \\&
  =\frac{1}{8}\left |\left | 
\sum _{n\in{\mathbb Z}' }a_n k_n e^{i\zben t} \right |\right |^2_{L^2(-T,T;K)}\,.   
\end{align*}
In the last equality we used we used $ -\beta_n=\zben $,  $ k_{-n} =k_n$ and $ a_n=a_{-n} $ (in the case of the sine sequence we have $ -a_n=a_{-n} $).

Inequalities~(\ref{eq:laDEFIrieszCONDbecSUFF}) hold by assumption for the last series, hence also for the first one.\zdia

This proof has been adapted from~\cite{Gubreeve}, where it is proved that the opposite implication is false.

Nothing that the transformation
\[ 
\sum \zaa_n e^{i\zben t}k_n\mapsto \sum \zaa_ne^{-i\zben T}e^{i\zben\tau}k_n\,:\qquad L^2(-T,T;K)\mapsto L^2(0,2T;K)
 \]
 is bounded and boundedly invertible (use $ |\mathcal{I}m\,\zben| $ bounded), we have also:
 \begin{Corollary}\ZLA{Corollary:daExpAcOS}
Let conditions~(\ref{eq:parita})
hold.  If the sequence $ \{ e^{i\zben t}k_n\}_{n\in\bbZ'} $ is a Riesz sequence in $ L^2( 0,2T,K) $, then the sequences
in~(\ref{eq:succeSINcosDAfareRiesz})
  are Riesz sequences in $ L^2(0,T;K) $. 
 \end{Corollary}

 Now we consider a Riesz \emph{basis} $ \{e_n\} $ in $ L^2(0,T;K) $ and a time $ T_0<T $. Then $ \{e_n\} $ is complete in $ L^2(0,T_0;K) $ but it is not a Riesz sequence since every element of $ L^2(0,T_0;K) $ has infinitely many representation as a series $ \sum a_n e_n $ (one such representation for every extension which belongs to $ L^2(0,T;K) $).
 
 Let $ {\mathbb J}_0 $ be the operator from $ L^2(0,T_0;K) $
 to $ l^2 $ given by
 \[ 
 {\mathbb J}_0f=\left \{\langle f,e_n\rangle_{L^2(0,T_0;K)}\right \}\,.
  \]
 We prove:
 \begin{Lemma}\ZLA{Lemma:delleSOTTOS}
The codimension of the range of the operator $  {\mathbb J}_0 $ is infinite.
\end{Lemma}
\zProof We know that 
\[
 \left ({\rm im}\,{\mathbb J}_0\right )^\perp=\ker {\mathbb J}_0^*  
 =\left \{
 \{c_n\}\in l^2\,:\quad \sum c_n e_n=0 \qquad {\rm in}\ L^2(0,T_0;K)\,.
 \right \}
 \]
 Every sequence $ \{c_n\} $ such that $ \sum c_n e_n=0 $ in $ L^2(0,T_0;K) $ while $ \sum c_n e_n\neq 0 $ in $ L^2(T_0,T ;K) $
 belongs to $ \ker {\mathbb J}_0^*   $, and conversely (a part the null element). 
 
 If  $ \ker {\mathbb J}_0^*   $ where finite dimensional then its image under the map
 \[ 
 \{c_n\}\mapsto \sum c_n e_n\,:\qquad l^2\mapsto L^2(0,T;K)
  \]
  would be finite dimensional, which is not true. 
  This ends the proof.\zdia

  \section{\ZLA{sect:Telegraph}Preliminaries on the telegraph equation}
  We called (generalized) telegraph equation the equation~(\ref{eq:telegrafisti}), obtained from Eq.~(\ref{eq:SisteMEMsecondordine}) when $ M(t)=0 $. Of course, Theorem~\ref{teo:regoSOLUZ} holds in particular for the telegraph equation
  and the definition of controllability can be applied to the telegraph equation too.

  Controllability, using \emph{real valued controls,} of the telegraph equation at a certain time   $ T  $  (and so also at larger times) has been proved in~\cite{Soriano} under standard assumption on $ \Gamma $. Here we don't  make any use of the explicit assumptions on $ \Gamma $, we just use controllability of the telegraph equation, but it is known that controllability of a wave type equation imposes conditions on the ``size'' of the active part $ \Gamma $ of $ \partial \ZOMq $, see~\cite{Bardos,TriggianiLIMIT}.

  A trivial observation is as follows: we can change at will the value of $ c $ without affecting controllability. In fact, $ u(x,t)=e^{bt}w(x,t) $ solves a telegraph equation like the one of $ w $, but with
  \[ 
  \mbox{$ c $, $q(x)$ replaced with $ c+b $ and $ q(x)-b^2 -cb$ }
   \]
   while $ a(x) $ remains unchanged. The eigenvalues depends on this transformation, but the asymptotic estimate and Lemma~\ref{eq:LemmaSUconveSERIE} are not affected.
   
   So, we might choose $ b=-c $ and cancel the velocity term but we shall see that a different choice is more usefull when studying controllability of the viscoelastic system.

  Controllability in time $ T $ is equivalent to surjectivity of the     map   $ f\,\mapsto (w,w_t) $ (acting from $ L^2(G_T) $ to $ L^2(\ZOMq)\times H^{-1}(\ZOMq) $). Using standard arguments, it is easy to see that this is equivalent to the following property:
  \begin{Lemma}\ZLA{LemmaHUMtelegr}
The telegraph equation is controllable in time $ T  $ iff   there exist $ m=m_T>0 $, $ M=M_T>0 $ such that the following inequality holds:
\begin{equation}
\ZLA{eq:dirINVEteleg}
m\left ( \|\phi_0\|^2_{H^1_0(\ZOMq)}+\|\phi_1\|^2_{L^2(\ZOMq)}\right )
\leq
\int _{G_T}\|\zg_a\phi\|^2\ZD G_T 
\leq M  \left ( \|\phi_0\|^2_{H^1_0(\ZOMq)}+\|\phi_1\|^2_{L^2(\ZOMq)}\right )\,.
\end{equation}
Here $ \phi $ denotes the solution of the adjoint system
\begin{equation}
\ZLA{eq:adjointTELEGRAPH}
\begin{split}
 \displaystyle
&  \phi _{tt}=2c \phi_t+\nabla\cdot\left (a(x)\nabla\phi\right )+q(x)\phi\,,\\
 \displaystyle
&  \phi(\cdot,0)=\phi_0(x)\in H _{0}^1(\ZOMq)\,,\ 
\phi_t(\cdot,0)=\phi_1(x)\in L^2(\ZOMq)\,,\ \phi _{| _{\partial\ZOMq}}=0\,.
\end{split}
\end{equation}
\end{Lemma}
The inequality from above holds   for every $ T $, even   when the system is non controllable. Controllability instead is crucial for the inequality from below.

Lemma~\ref{LemmaHUMtelegr} is the starting point of the HUM method (see~\cite{LionsLIBRO}). The proof in~\cite{Kim2} essentially extends similar inequalities to the case that the kernel is not zero.

Now we list three consequences which we shall use below:

\begin{Lemma}\ZLA{Corollary:primosulbordo}
Let $ \Gamma $ be so chosen that controllability holds for the telegraph equation at some time $ T $.
Let $ A\phi=0 $. If $ \zg_a\phi=0 $ then we have $ \phi=0 $.
\end{Lemma}
\zProof In fact, $ \phi(x,t)= e^{ 2c  t}\phi(x) $ solves~(\ref{eq:adjointTELEGRAPH}) with $ \phi(x,0)=\phi(x) $ and $ \phi_t(x,0)= 2c \phi(x) $.   The choice of $ \Gamma $
 implies that the left inequality in~(\ref{eq:dirINVEteleg}) holds for $ T $ sufficiently large, and the integral is zero. Hence $ \phi=0 $.\zdia

 We can extend Lemma~\ref{Corollary:primosulbordo} to nonzero eigenvalues:
 \begin{Lemma}\ZLA{lemma:perTRACCIAcombi}
 Let $ \Gamma $ be so chosen that controllability holds for the telegraph equation at some time $ T $.
 Let $ A\phi =\n\phi   $.
  If $ \zg_a\phi=0 $ then $ \phi =0 $.
 \end{Lemma}
 \zProof
 For the proof, we apply the left inequality in~(\ref{eq:dirINVEteleg}) (which holds for $ T $ sufficiently large).
 Let
 \begin{equation}
\ZLA{eq:DefinitionZBEN}
\zben=\sqrt{\zl_n^2-c^2}\,.
\end{equation}
  We distinguish the case $ \zben\neq 0 $ and the case $ \zben=0 $. In the first case, the inequality is applied to the solution  $ \phi(x,t)=e^{ct}\phi(x)\sin\zben t $. Otherwisewe use the solution $ \phi(x,t)=e^{ct}\phi(x)\cos\zben t $.\zdia

   We recall that a sequence $ \{\psi_n\} $ in a Hilbert space is \emph{almost normalized} (or \emph{almost normal})
when there exist  $ m>0 $ and $ M $ such that
\[ 
m\leq \|\psi_n\|\leq M\,.
 \]
 \begin{Lemma}\ZLA{lemma:condiSUGAMMa}
If $ \Gamma $ is so chosen that controllability holds for the telegraph equation at time $ T $, then the sequence $ \{(\zg_a\phi_n)/\zl_n\} $ is almost normalized in $ L^2(\Gamma) $ (here $ \phi_n $ are the normalized eigenfunctions af $ A $ whose eigenvalue is not zero).
\end{Lemma}
\zProof We solve Eq.~(\ref{eq:adjointTELEGRAPH}) with initial conditions
\[ 
\phi (x,0)=0\,,\qquad \phi_t(x,0)=\phi_n(x)\,.
 \]
 The solution is ($ \zben $ is defined in~(\ref{eq:DefinitionZBEN}))
 \[ 
 \phi (x,t)=\frac{1}{\zben}e^{ct}\phi_n(x)\sin\zben t 
  \]
  (we might have $ \zben=0 $ for a  {\em finite \/}  set of indices. We disregard these 
  elements without affecting the result). Using $ \|\phi_n\|_{L^2(\ZOMq) }=1 $, inequality~(\ref{eq:dirINVEteleg}) gives
  \[ 
  m\leq \int _{G_T}\left (
  \frac{\zl_n}{\zben} e^{ct}\sin \zben t
  \right )^2\left |
  \frac{\zg_a\phi_n}{\zl_n}
  \right |^2\ZD G_T<M\,.
   \]
  The result follows since, when $ c\neq 0 $,
  \[ 
  \lim _{n\to+\ZIN } \frac{\zl_n}{\zben}=1\,,\qquad  \lim _{n\to+\ZIN }\int_0^T e^{2ct}\sin^2\zben t\ZD t=\frac{1}{4c}\left (e^{2cT } -1\right )\,. 
   \]
 Instead, if $ c=0 $ 
  \[ 
   \lim _{n\to+\ZIN }\int_0^T  \sin^2\zben t\ZD t=\frac{1}{2}T\,.\zdiaform
   \]

 See~\cite{TAOcorrect} for the idea of this proof.

%
%
%
%
%
 \subsection{Moment method for the telegraph equation}
 
 The following computations make sense for smooth controls and are then extended to square integrable controls by continuity. Let
 \[ 
 w_n(t)=\int _{\ZOMq} w(x,t)\phi_n(x)\ZD x
  \]
 ($ \{\phi_n\} $ is the orthonormal basis of $ L^2(\ZOMq) $, of   eigenvectors of the operator $ A  $ in~(\ref{eq:DefiOPeratore-A})). Then, $ w_n(t) $ solves
 \[ 
 w_n''=2cw_n'-\zl_n^2 w_n -\int _\Gamma (\zg_a \phi_n)f(x,t)\ZD\Gamma\,.
  \]
  So, with
  \[ 
  \zben=\sqrt{\zl_n^2-c^2}\,,
   \]
   we have
   \begin{equation}\ZLA{eq:diWnPerTELEnobenZERO}
   w_n(t)=-\int _{G_t} e^{cs}\left [
   \frac{\zg_a\phi_n}{\zben} \sin\zben s
   \right ]f(x,t-s)\ZD G_t
    \end{equation}
    if $ \zben\neq 0 $. We might have $ \zben=0 $ for a finite number of indices and in this case~(\ref{eq:diWnPerTELEnobenZERO}) has to be replaced with
 
   \begin{equation}\ZLA{eq:diWnPerTELEuqualZERO}  
   w_n(t)= -\int _{G_t} s e^{cs}\left [
   \zg_a\phi_n
   \right ] f(x,t-s)\ZD G_t\,.
   \end{equation} 
   So, we have
   \begin{align}
\ZLA{eq:teleSERIEdiDEFORM}
&  -w(x,t) 
=\sum \phi_n(x) 
\int _{G_t} e^{cs}\left [
   \frac{\zg_a\phi_n}{\zben} \sin\zben s
   \right ]f(x,t-s)\ZD G_t
 \,,\\
 \ZLA{eq:teleSERIEdiVELOC}
 &  -w_t(x,t) 
=\sum \zben\phi_n(x)\int _{G_t}e^{cs}\frac{\zg_a\phi_n}{\zben}\left [
\frac{c}{\zben}\sin\zben s+\cos\zben s
\right ]f(x,t-s)\ZD G_t\,.
\end{align}
If $ \zben=0 $ then the corresponding term in~(\ref{eq:teleSERIEdiDEFORM})
is replaced with~(\ref{eq:diWnPerTELEuqualZERO})  while in~(\ref{eq:teleSERIEdiVELOC}) it is replaced with
\begin{equation}\ZLA{eq:MOmePROtelegEQUA2zbenequalZERO}
\int _{G_t} (1+cs)e^{cs}\left (\zg_a \phi_n\right ) f(x,t-s)\ZD G_t\,.
 \end{equation} 
    
    For any $ k>0 $   such that $ kI-A $ is positive,  
 the sequence $   \{ \phi_n (\sqrt{k+\zl_n^2} )^{-1}  
  \} $    is an orthonormal basis of $ \left ({\rm dom}\, (kI-A)^{1/2}\right ) $ and so $ \{  \phi_n  \sqrt{k+\zl_n^2}  \} $   is an orthonormal basis of $ \left ({\rm dom}\, (kI-A)^{1/2}\right )' $. This space is unitary equivalent to $ H^{-1}(\ZOMq) $ since (from~\cite[Theorem~1-D]{Fujuwara})
    \[ 
     \left ({\rm dom}\, (kI-A)^{1/2}\right )=H^1_0(\ZOMq)\,.
     \]
     Hence, every $ \chi \in H^{-1}(\ZOMq) $ is represented as
     \[ 
     \chi=\sum \chi_n \left (\sqrt{k+\zl_n^2}\right )\phi_n
      \]
      for an arbitrary $ \{\chi_n\}\in l^2 $. Then we have also
     \[ 
     \chi=\sum_{\zben =0} \left ( \chi_n  \sqrt{k+c}\right )\phi_n+
    \sum_{\zben\neq 0} \left (\chi_n \frac{\sqrt{k+\zl_n^2} }{\zben}\right )\left (\zben \phi_n\right )\,.
      \]
      It follows that the sequence $ \{\zben \phi_n\} $ is a Riesz basis of $ H^{-1}(\ZOMq) $ (in this sequence the term $ \zben\phi_n $ has to be replaced with $ \phi_n $ if $ \zben=0 $).

     Equalities~(\ref{eq:teleSERIEdiDEFORM})--(\ref{eq:teleSERIEdiVELOC}) show that controllability at time $ T $ of the telegraph equation is equivalent to solvability of the following moment problem
 \begin{align}
\ZLA{eq:MOmePROtelegEQUA1}
&\int _{G_T} e^{cs}\left [
   \frac{\zg_a\phi_n}{\zben} \sin\zben s
   \right ]f(x,T -s)\ZD G_T=\xi_n\\
\ZLA{eq:MOmePROtelegEQUA2}
&\int _{G_T}e^{cs}\frac{\zg_a\phi_n}{\zben}\left [
\frac{c}{\zben}\sin\zben s+\cos\zben s
\right ]f(x,T-s)\ZD G_T
=\eta_n
\end{align}
where $ \{\xi_n\} $ and $ \{\eta_n\} $ belong to $ l^2 $.

We noted that when $ \zben=0 $  the corresponding terms in~(\ref{eq:MOmePROtelegEQUA1}) and~(\ref{eq:MOmePROtelegEQUA2}) have to be replaced   respectively with~(\ref{eq:diWnPerTELEuqualZERO}) or~(\ref{eq:MOmePROtelegEQUA2zbenequalZERO}).
 In order to have a unified formulation, we introduce
 \[ 
 \mathcal{J}=\{n\,:\ \zben=0\}
  \]
  (a finite set of indices) and
 $ c_n=\eta_n+i\xi_n $. Then, $ \{c_n\} $ is an arbitrary (complex valued) $
  l^2 $ sequence. The moment problem~(\ref{eq:MOmePROtelegEQUA1})-(\ref{eq:MOmePROtelegEQUA2}) reduces to the following:  
\begin{equation}\ZLA{eq:momentiTELEGRcomplesso}
 \begin{array}{l}
\displaystyle 
 \int_{G_T}\left (\frac{\zg_a\phi_n}{\zben}\right )\left [ e^{i\zben s}+\frac{c}{\zben}\sin\zben s\right ] \left (e^{cs}f(x,T-s)\right )\ZD G_T=c_n\,,\quad  n\notin \mathcal{J }\\
 \displaystyle 
 \int _{G_T} (1+cs+is) \left (\zg_a\phi_n\right ) \left (e^{cs}f(x,T-s)\right )\ZD G_T=c_n\,,\quad  n\in \mathcal{J } 
 \end{array}
\end{equation}
where $ f $ is \emph{real valued.}
 Now we introduce, for $ n<0 $,
 \[ 
 \zben=-(\beta_{-n})\,,\quad \phi _{n}=\phi_{-n}\,,\quad  \zl_n= \zl_{-n}\,.
  \]
   Then we consider the moment problem~(\ref{eq:momentiTELEGRcomplesso}) with
   \[ 
   n\in \mathbb{Z}'=\mathbb{Z}-\{0\}\,,\quad \{c_n\}\in l_2(\mathbb{Z}') 
    \] 
   with \emph{complex valued} controls.
    The next formula proves that the moment problem in $  \mathbb{Z}' $ and \emph{complex}  control $ f $ is solvable if and only if the moment  problem~(\ref{eq:momentiTELEGRcomplesso}) is solvable in
   $ l^2(\mathbb N) $, with \emph{real valued control.} In fact, let  
   $ f(s)=h(s)+ik(s) $ and put (for $ n\notin\mathcal J $ . Obvious modification for $ n\in \mathcal J  $)
    \begin{eqnarray*}
 &&   \int _{G_T}\frac{\zg_a\phi_n}{\zben}\left \{ \left (\cos\zben s+\frac{c}{\zben}\sin\zben s\right )h(s)\right.\\
&& \left.-(\sin\zben s)k(s)+i\left [\left(\cos\zben s+\frac{c}{\zben}\sin\zben s\right )k(s)+h(s)\sin\zben s\right ]\right \}\ZD G_T\\
&& =\omega_n+i\nu_n=A_n+B_n+iC_n+iD_n
     \end{eqnarray*}
     
     We sum the terms with $ n>0 $ with the corresponding terms with $ -n $ of the \emph{conjugate equality.} Using the parity of the functions we find the moment problem~(\ref{eq:momentiTELEGRcomplesso}) for $ n>0 $, $ f=h/2  $ and the right hand side
     \[ 
     (\omega_n +\omega_{-n})+i(\nu_n- \nu_{-n})\,,
      \]
      which can be arbitrarily assigned.
       Conversely, let us assume that the moment problem~(\ref{eq:momentiTELEGRcomplesso}) is solvable for $ n>0 $, and real valued control and consider the previous equalities for $ n\in\mathbb{Z}' $.      
The sequences $\{ A_n\} $, $\{B_n\}$, $\{C_n\}$, $\{D_n\}$ can 
  be arbitrarily assigned in $ l^2 (\mathbb{Z}')$ when the  moment problem~(\ref{eq:momentiTELEGRcomplesso})     is solvable, with the conditions
    \[ 
    A_n=A _{-n}
\,,\qquad B_n= -B _{-n}\,,\qquad C_n=C _{-n}\,,\qquad D_n=- D _{-n}\,.     \]
    Now we fix $ \{\eta_n+i\xi_n\} $ in $ l^2(\bbZ') $ and we see that the equalities
 \[
 \begin{split}
  &   A_n+B_n+iC_n+iD_n=\eta_n+i\xi_n\,, \\
 &  A_n-B_n+iC_n-iD_n=\eta _{-n}+i\xi _{-n}\,, 
 \end{split}
 \qquad n>0 
  \]
     can be realized. Hence, our assumption that the telegraph equation is solvable in time $ T $ \emph{with real valued controls} can be rephrased as follows:
     \begin{Theorem}
     The telegraph equation is controllable in time $ T $
if and only if   the moment problem~(\ref{eq:momentiTELEGRcomplesso}) is solvable for every complex sequence $ \{c_n\}\in l^2(\bbZ') $,  with \emph{complex valued} control functions $ f\in L^2(G_T) $.
\end{Theorem}
 It has an interest to note that when $ \{\xi_n\} $ and $ \{\eta_n\} $ are arbitrary in $ l^2 $, the same holds for the sequence $ \{\eta_n+i\xi_n-(\zg/\zben)\xi_n\} $,  for every number $ \zg $. So, we have also:
 \begin{Theorem}
  The telegraph equation is controllable in time $ T  $ if and only if
  the moment problem~(\ref{eq:momentiTELEGRcomplesso}) with $ c $ replaced with any number $ \zg $ (in particular with $ \zg=0 $) is solvable for every complex sequence $ \{c_n\}\in l^2(\bbZ') $.
\end{Theorem}
Using~\cite[p.~34]{AvdoninIVANOV} we then get:
\begin{Theorem}\ZLA{teo:RieszPERtele}
Let us consider the sequence whose elements are
\[ 
\left (e^{i\zben t}+\frac{c\zg}{\zben}\sin\zben t\right )\frac{\zg_a\phi_n}{\zben}
 \quad  n\notin \mathcal{J}\,,\qquad (1+(i+\zg)s)\zg_a\phi_n  \quad n\in \mathcal{J}
 \]
 where $ \zg $ is any fixed complex number. If the telegraph equation is controllable in time $ T $, then    this sequence is a Riesz sequence in $ L^2(G_T) =L^2(0,T;L^2(\Gamma))$  and conversely.
\end{Theorem}
 
   \section{Moment problem and controllability of viscoelastic systems}
   
   The computations are simplified if we perform first a transformation introduced in~\cite{PandIEOT}. This transformation is  more transparent if we integrate both the sides of~(\ref{eq:SisteMEMsecondordine}). We recall that we can assume zero initial conditions (and affine term) without restriction so that we get
   \[
w_t(t)=2cw(t)+\intt \tilde  N(t-s)\left (\nabla\cdot\left (a(x)\nabla w(s)\right )+q(x)w(s)\right )\ZD s
\] 
 with conditions
 \[ 
 w(0)=0\,,\qquad w_{|_{\Gamma}}(t)=f(t)\,,\quad  w_{|_{\partial\ZOMq\setminus\Gamma}}(t)=0
  \]
 and
 \[ 
\tilde N(t)=1+\intt M(s)\ZD s\,.
  \]
 We introduce
 \[ 
 \zthe(x,t)=e^{2\zg t}w(x,t)\,,\qquad  \zg =-M(0)/2=-\tilde N'(0)/2\,.
  \]
 We see that $ \zthe $ solves the following equation, where
 \[ 
 \zaa=c+\zg\,: 
  \]
 
\begin{equation}
\ZLA{eq:viscoPRIMORD}
\zthe_t =2\zaa \zthe(t)+\intt   N(t-s)\left (\nabla\cdot\left (a(x)\nabla \zthe(s)\right )+q(x)\zthe(s)\right )\ZD s
\end{equation}
 and
 \[ 
 \zthe(0)=0\qquad \zthe_{|_{\Gamma}}(t)=e^{2\zg t}f(t)\,,\quad \zthe_{|_{\partial\ZOMq\setminus\Gamma}}(t)=0
  \]
 (the functions $ e^{2\zg t}f(t) $ will be renamed $ f(t) $)  and
 \[ 
 N(t)=e^{2\zg t}\tilde N(t)\quad \mbox{
 so that $ N(0)=1 $ and $ N'(0)=0 $ 
 } \,.
  \]
 The condition $ N'(0)=0 $ simplifies the following computations.  
 
 Noting that
 \[ 
 w_t=e^{-2\zg t}\left (\zthe_t-2\zg \zthe\right )\,,
  \]
 controllability of the pair $ (w,w_t) $   is equivalent to controllability of the pair  $(\zthe,\zthe_t) $. So, from now on we study the controllability of the pairs $ (\zthe(t),\zthe_t(t)) $ where $ \zthe  $ solves Eq.~(\ref{eq:viscoPRIMORD}).
 
 \begin{Remark}
{\rm
Computing the derivative of both the sides of Eq.~(\ref{eq:viscoPRIMORD}) we get
\[ 
\zthe _{tt}=2\zaa\zthe_t+
\nabla\cdot\left (a(x)\nabla \zthe \right )+q(x)\zthe 
+\intt   N(t-s)\left (\nabla\cdot\left (a(x)\nabla \zthe(s)\right )+q(x)\zthe(s)\right )\ZD s\,.
 \]
 The telegraph equation which corresponds to this system is
\begin{equation}
\ZLA{eq:TeleCONFRO}
\zthe _{tt}=2\zaa\zthe_t+
\nabla\cdot\left (a(x)\nabla \zthe \right )+q(x)\zthe 
\end{equation}
 We shall prove controllability of the viscoelastic system comparing it with the telegraph equation~(\ref{eq:TeleCONFRO}).\zdia
}
\end{Remark}

 We project $ \zthe( t) $ along the eigenvector $ \phi_n $. Let
 \[ 
 \zthe_n(t)=\int _{\ZOMq}\zthe(x,t)\phi_n(x)\ZD x 
  \]
so that
  \[ 
  \zthe_n' =2\zaa\zthe_n-\zl_n^2 \intt N(t-s)\zthe_n(s)\ZD s-\intt N(t-s) \left [
  \int _{\Gamma} (\zg_a\phi_n) f(x,s)\ZD \Gamma
  \right ]\ZD s\,.
   \]
   For every $ n $ we introduce the functions $ z_n(t) $ which solve
   \begin{equation}
\ZLA{eq:equazDIzMINUSCOLO}
z_n'=2\zaa z _n-\zl_n^2 \intt N(t-s)z _n(s)\ZD s\,,\qquad z_n(0)=1\,.
\end{equation}
Then we have:
\begin{align}
\nonumber&& \zthe_n(t)=-\intt z_n(\zt)\int_0^{t-\zt}N(t-\zt-s)\int _{\Gamma} (\zg_a\phi_n) f(x,s)\ZD\Gamma\, \ZD s\,\ZD \zt\\
\ZLA{eq:diTHETAn}&&= -\int _{G_t}\left \{
\int_0^s N(s-\zt)z_n(\zt)\ZD\zt
\right\} (\zg_a\phi_n) f(x,t-s)\ZD G_t \\
\ZLA{eq:diTHETAprimoN}&&\zthe_n'(t)=-\int _{G_t}\left [
z_n(s)+\int_0^s N'(s-\zt) z_n(\zt)\ZD \zt
\right ](\zg_a\phi_n) f(x,t-s)\ZD G_t\,.
\end{align}
Note that these computations are justified thanks to Theorem~\ref{teo:regoSOLUZ} (see also~\cite[Appendix]{PandAMO}), which justify also the following equalities, respectively in $ L^2(\ZOMq) $ and $ H^{-1}(\ZOMq) $:
\[ 
\zthe(t)=\sum _{n=1}^{+\ZIN} \zthe_n(t) \phi_n(x)\,,\qquad \zthe_t(t)=\sum _{n=1}^{+\ZIN} \zthe_n'(t)\phi_n(x)\,.
 \]
Let $ \{\xi_n\}\in l^2 $ and $ \{\eta_n\} \in l^2 $ be the sequence of the coefficients of the expansions of the targets $ \xi $ and $ \eta $ in series of, respectively, $ \{\phi_n\} $ and $ \{\zben\phi_n\} $ ($\zben\phi_n$  replaced with  $ \phi_n  $ if $ \zben=0 $). We see that controllability at time $ T $ is equivalent to the solvability of the following moment problem:
\begin{equation}
\ZLA{eq:MomePROcomplPERvisco}
\begin{array}{l}
\displaystyle \int _{G_T} Z_n(t)\frac{\zg_a \phi_n}{\zben} f(x,T-s)\ZD G_T=c_n=-(\eta_n+i\xi_n)\,,\qquad n\notin \mathcal{J}\\
\displaystyle \int _{G_T} Z_n(t) (\zg_a \phi_n  ) f(x,T-s)\ZD G_T=c_n=-(\eta_n+i\xi_n)\,,\qquad n\in \mathcal{J}
\end{array}
 \end{equation}  
  where, for $ n>0 $,
\begin{equation}
\ZLA{eq:equazdiZMaiuSCOLO}
\displaystyle 
Z_n(t)=
\left\{
\begin{array}{ll}
\displaystyle z_n(t)+\intt N'(t-s) z_n(s)\ZD s+i\zben \intt N(t-s)z_n(s)\ZD s\,, & n\notin \mathcal{J}\,,\\
\displaystyle z_n(t)+\intt N'(t-s)z_n(s)\ZD s+i\intt N(t-s) z_n(s)\ZD s\,,
& n\in \mathcal{J}
\end{array}\right.
\end{equation} 
   (we recall that if $  n\in \mathcal{J} $ then the element $ \zben\phi_n $ of the basis of $ H^{-1}(\ZOMq) $ has to be replaced with $ \phi_n $).
   
 It is convenient to reformulate the moment problem with $ n\in\bbZ'$. This is done using the following definitions:
 
   \[
   z_n(t)=z _{-n}(t)\,,\quad \phi_n(x)=\phi _{-n}(x)\,,\qquad \beta _{-n}=- \zben\,,\qquad \zl_{-n}=\zl_n\,,\qquad n\notin\mathcal{J}\,.
    \]
    These inequalities imply
    \[ 
    Z_{-n}(t)=\overline{Z}_{n}(t)  \,,\qquad n\notin\mathcal{J}\,.
     \]
   We impose an analogous condition also if $ n\in\mathcal{J} $:
    \[ 
    Z_{-n}(t)=\overline{Z}_{n}(t) \qquad\mbox{ if $ n\in\mathcal{J} $.}
     \] 
     So, we can consider the moment problem~(\ref{eq:MomePROcomplPERvisco}) with $ n\in\bbZ'=\bbZ\setminus\{0\} $,  extending the definition of  $  \mathcal{J} $ as the set of positive or negative indices for which $ \zben =0$.
     
       In order to prove Theorem~\ref{teo:controllPRINCIP},  we must prove:
       \begin{Theorem}\ZLA{teo:princiMAIN}
Let the telegraph equation~(\ref{eq:TeleCONFRO}) be controllable at time $ T $.   Let, for  $ n\in\bbZ' $,
\begin{equation}\ZLA{eq:DefiPSIn}
\Psi_n=\frac{\zg_a\phi_n}{\zben} \quad n\notin\mathcal{J}\,,\qquad \zg_a\Phi_n\quad n\in\mathcal{J}\,.
\end{equation}
Then, the sequence $ \{Z_n(t)\Psi_n\} $ is a Riesz sequence in $ L^2(G_T)=L^2(0,T;L^2(\Gamma)) $.
\end{Theorem}
The remaining part of this section is devoted to the proof of theorem~\ref{teo:princiMAIN}.
 
\subsection{The functions $ Z_n(t) $}

Let
\[ 
K_n(t)=N'(t)+i\zben N(t)\quad {\rm if}\quad n\notin \mathcal{J}\,,\qquad K_n(t)=N'(t)+iN(t)\quad {\rm if}\quad n\in\mathcal{J}\,.
 \]
 The right hand side of the equality~(\ref{eq:equazdiZMaiuSCOLO}) is a variation of constants formula, so that  (compare~(\ref{eq:equazDIzMINUSCOLO})) $ Z_n(t) $ solves
 \begin{equation}
\ZLA{eq:equadiff-di-Zn-PRIMORD}
Z_n'=2\zaa Z_n-\zl_n^2\intt N(t-s) Z_n(s)\ZD s+K_n(t)\,,\qquad Z_n(0)=1\,.
\end{equation}
   Hence also
   \begin{equation}
\ZLA{eq:equadiff-di-Zn-SECORD}
\begin{array}{l}
\displaystyle   Z_n''=2\zaa Z_n'-\zl_n^2Z_n-\zl_n^2\intt N'(t-s)Z_n(s)\ZD s+K_n'(t)\,,\\[4mm]
\left\{\begin{array}{l}
\displaystyle  Z_n(0) = 1\,,\\[2mm]
\displaystyle  Z_n'(0)=
 2\zaa+i\zben \ \mbox{($n\notin{\mathcal J}$)}\,,\quad 
 Z_n'(0)=
 2\zaa+i \  \mbox{($ n\in{\mathcal J}$)}\,.
%
%
\end{array}\right.
\end{array}
\end{equation}
Then we have the following representation formulas, where now
\[ 
\zben=\sqrt{\n^2-\zaa^2}
 \]
\begin{align*}
 & \mbox{if $ n\notin\mathcal{J}$ then}  \\& Z_n(t)=e^{\zaa t}e^{i\zben t}+e^{\zaa t}\frac{\zaa}{\zben}\sin\zben t\\
&   +\frac{1}{\zben}\intt e^{\zaa(t-s)}\sin\zben(t-s)\left [
K_n'(s)-\zl_n^2\ints N'(s-r)Z_n(r)\ZD r
\right ]\ZD s \,,\\
 &\mbox{if $n\in\mathcal {J}$ then }\\
 &   Z_n(t)=e^{\zaa t}\left (1+(\zaa+i) t\right )\\
 & +\intt e^{\zaa(t-s)}(t-s)\left [\left (N''(s)+iN'(s)\right )-\zaa^2\intr N'(r-s)Z_n(s)\ZD s\right ]\ZD r\,.
\end{align*}
We introduce
\[ 
S_n(t)=e^{-\zaa t}Z_n(t)
 \]
and we see that, for $ n\notin{\cal J} $,
\[S_n(t)=G_n(t)-\frac{\zl_n^2}{\zben}\intt \sin\zben(t-s)\ints \left ( e^{-\zaa(s-r)}N'(s-r)\right ) S_n(r)\ZD r\,\ZD s
\]
where 
\begin{align}
\nonumber&G_n(t)=e^{i\zben t}+\frac{\zaa}{\zben} \sin\zben t+\frac{1}{\zben} \intt e^{-\zaa s}\left [N''(s)+i\zben N'(s)\right ] \sin\zben(t-s)\ZD s\\
\nonumber& e^{i\zben t}+\frac{\zaa-N'(0)}{\zben}\sin\zben t+\intt N'(t-s)e^{-\zaa(t-s)} \left [ e^{i\zben(t-s)} +\frac{\zaa}{\zben}\sin\zben(t-s)\right ]\ZD s
\\
&\ZLA{eq:DEFIdiGn}
=e^{i\zben t}+\frac{\zaa}{\zben}\sin\zben t+\intt N'(t-s)e^{-\zaa (t-s)} \left (e^{i\zben s}+\frac{\zaa}{\zben}\sin\zben s\right )\ZD s 
\end{align}
(in the last step we used $ N'(0)=0 $).

Instead, for $ n\in{\cal J} $ we  have
\[ 
G_n(t)=1+(\zaa+i)t+\intt e^{-\zaa(t-s)}N'(t-s)\left [1+(\zaa+i)s\right ]\ZD s\,.
 \]
Using the fact that the linear transformation
\[  y\mapsto y(t)+\intt e^{-\zaa(t-s)}N(t-s)y(s)\ZD s  \]
is bounded with bounded inverse, we get, using~Theorem~\ref{teo:RieszPERtele}:
\begin{Theorem}
Let   the telegraph equation~(\ref{eq:TeleCONFRO}) be controllable at time $ T $. The, the sequence $ \{G_n(t)\Psi_n\} $ is Riesz in $ L^2(G_T) $.
\end{Theorem}

We shall  need asymptotic estimates of $ S_n(t) $ which holds for large $ n $. So, when deriving these estimates, we can work with $ n\notin{\cal J} $.

We introduce the notations
\[ 
N_1(t)=e^{-\zaa t}N'(t) \quad\mbox{so that $ N_1(0)=0 $,}\qquad
 \mu_n=\frac{\zl_n^2}{\zben^2}\,.
 \]
  Then, an integration by parts gives
  \begin{eqnarray}
\nonumber &&
S_n(t)=G_n(t)-\mu_n\intt N_1(t-r)S_n(r)\ZD r\\
\ZLA{Eq:equaIntegrperSNpriFORM}&& 
+\mu_n\intt\left (\int_0^{t-r}N_1'(t-r-s)\cos\zben s\ZD s\right ) S_n(r)\ZD r\,.
\end{eqnarray}
Gronwall inequality shows:
\begin{Lemma}
For every $ T>0 $ there exists $ M=M_T $ such that for every $ t\in[0,T] $ and every $ n $ we have:
\[ 
|S_n(t)|\leq M
 \]
\end{Lemma}

We integrate by parts again the last integral in~(\ref{Eq:equaIntegrperSNpriFORM}) and we get
\begin{align}
\nonumber   S_n(t)&=G_n(t)-\mu_n\intt N_1(t-r) S_n(r)\ZD r  
 %
\\
  \ZLA{Eq:equaIntegrperSN-FORMsemi-finale}&+\frac{\mu_n}{\zben}
\intt\left (
N_1'(0)\sin\zben(t-r) +\int_0^{t-r}N_1''(t-r-s)\sin\zben s\ZD s
\right )S_n(r)\ZD r\,. 
\end{align}

We note that
\[ 
1-\mu_n=-\frac{\zaa_n}{\zben^2}
 \]
and we rewite the previous equality as
\begin{align}
\nonumber &(S_n-E_n)+N_1\star (S_n-E_n)= \frac{N_1'(0) }{\zben}\intt  \sin\zben(t-r)S_n(r)\ZD r\\
\ZLA{eq:rintREviNUvers}& +\frac{1}{\zben}\intt 
N_1''(s)\int_0^{t-s}\sin\zben(t-s-r)S_n(r)\ZD r\,\ZD s+\frac{1}{\zben^2}M_n(t)\,.
\end{align}
Here
\begin{align*}
 &M_n(t) =- \zaa^2  \intt N_1(t-r) S_n(r)\ZD r
\\
&+\frac{\zaa}{\zben}\intt \left [N_1'(0)\sin\zben(t-r)+\int_0^{t-r}N_1''(t-r-s)\sin\zben s\ZD s\right ]\ZD r\,.
 \end{align*}
Using the definition of $ E_n(t) $ we see that
\begin{Theorem}
There exists a sequence $ \{M_n(t)\} $ of continuous functions defined for $ t\geq 0 $, bounded on bounded intervals and such that
\begin{equation}\ZLA{rappreEXPdiSn}
S_n(t)=e^{i\zben t}+\frac{M_n(t)}{\zben}\,.
 \end{equation}
 \end{Theorem}

Now we compute:
 \begin{align}
\nonumber& \intt S_n(r)\sin\zben (t-r)\ZD r
=\intt \left ( e^{i\zben r}+\frac{M_n(r)}{\zben}\right )\sin\zben(t-r)\ZD r
\\  
\ZLA{eq:SncontroSENO}&=-\frac{i}{2}t e^{i\zben t} +\frac{i}{2\zben} \sin\zben t 
 +\frac{1}{\zben} \intt M_n(r)\sin\zben(t-r)\ZD r\,.
\end{align}
We observe
\[
\frac{i}{\zben} t e^{i\zben t}=-\intt s e^{i\zben s}\ZD s+\frac{1}{\zben^2}\left (e^{i\zben t}-1\right ) 
 =-\intt sE_n(s)\ZD s+\frac{1}{\zben^2}M_n(t)\,.
\]
i.e.
\begin{align*}
&\frac{1}{\zben}\intt S_n(r)\sin\zben(t-r)\ZD r=-\frac{1}{2}\frac{it}{\zben}e^{i\zben t}+\frac{1}{\zben^2}M_n(t) 
 \\
 &=\frac{1}{2}\intt sE_n(s)\ZD s+\frac{1}{\zben^2}M_n(t)\,.
\end{align*}
 
We replace this expression in~(\ref{eq:rintREviNUvers}) and we rewrite the equality as
 
\begin{align*}
& (S_n-E_n)+N_1\star(S_n-E_n) 
\\&
=\frac{N_1'(0)}{2}\intt s E_n(s)\ZD s+\frac{1}{2}\intt N_1''(s)\int_0^{t-s}r E_n(r)\ZD r\,\ZD s+\frac{1}{\zben^2}M_n(t)
\\
&=\frac{1}{2}\intt N_1'(t-r)r E_n(r)\ZD r+\frac{1}{\zben^2}M_n(t)
\end{align*}
(as usual, the  functions $ M_n(t) $ are not the same at every step).

Let $ L(t) $ be the resolvent kernel of $ N_1(t) $ so that $ L(0)=0 $ and $ L(t) $ is twice differentiable. We have
\begin{align*}
&S_n(t)=E_n(t)+\frac{1}{2}\intt N_1'(t-s) s E_n(s)\ZD s\\
&-\frac{1}{2}\intt (sE_n(s))\left [\int_0^{t-s}L(t-s-r) N_1'(r)\ZD r\right ]\ZD s+\frac{1}{\zben^2}M_n(t)\,.
\end{align*}
In conclusion,
\begin{align} 
\nonumber&
\Psi_n S_n(t)=\Psi_nE_n(t)+\\
\ZLA{eq:formFinPartaAGG}&
\frac{1}{2}\intt s\left [  
N_1'(t-s)  - \int_0^{t-s}L(t-s-r)  N_1'(r)\ZD r
 \right ]\Psi_nE_n(s)\ZD s+\frac{1}{\zben^2}M_n(t) 
\end{align}
(note that we can replace $ \Psi_nM_n(t) $ with $ M_n(t) $ since $ \{\Psi_n\}$ is bounded in $ L^2(\Gamma) $).
The sequence whose elements are 
\[
\Psi_nE_n(t)+ 
\frac{1}{2}\intt s\left [ 
N_1'(t-s)  -\int_0^{t-s}L(t-s-r)  N_1'(r)\ZD r
\right ]\Psi_nE_n(s)\ZD s
\]
is the image of a Riesz sequence of $ L^2(G_T) $   under a linear bounded and boundedly invertible transformation. Hence, it is a Riesz sequence too so that, 
using Theorem~\ref{teo:PaleyWIENERbari} and Lemma~\ref{eq:LemmaSUconveSERIE}, we get:

\begin{Theorem}\ZLA{Lemma:RieszPERnGRANDE}
 Let the telegraph equation~(\ref{eq:TeleCONFRO}) be controllable at time $ T $. The following   hold:
\begin{itemize}
\item There exists $ N $ such that $ \{S_n(t)\Psi_n\} _{|n|>N} $ is a Riesz sequence in $ L^2(G_T) $;
\item the sequence $ \{S_n(t)\Psi_n\}_{n\in \bbZ'  } $  is a Riesz sequence in $ L^2(G_T) $  if and only if we can prove that it is $ \ZOM $-independent.
\end{itemize}
\end{Theorem}
Our final goal is the proof that $ \{S_n(t)\Psi_n\}_{n\in \bbZ'  }$
is $ \ZOM $-independent in $ L^2(G_T) $, which will finish the proof of
Theorem~\ref{teo:controllPRINCIP}.

\subsection{$ \ZOM $-independence}
In this section we prove that the sequence $ \{ S_n(t)\Psi_n\} $ (hence also $ \{ Z_n(t)\Psi_n\} $) is $ \ZOM $-independent in $ L^2(G_T) $ when the telegraph equation is controllable at time $ T $. We consider the equality
\begin{equation}\ZLA{eq:laNONomegaINDIP} 
\zSn \zaa_n S_n(t)\Psi_n=0 \qquad \mbox{in $ L^2(G_T) $}
 \end{equation}
 and we show that $ \{\zaa_n\}=0 $. Theorem~\ref{Lemma:RieszPERnGRANDE}  implies that $ \{\zaa_n\}\in l^2 $ and we proceed in several steps, whose key points are:
 \begin{itemize}
 
\item the sequence $ \{\zaa_n\} $ is ``regular''. In particular, $ \zaa_n=\zg_n/\zben^3 $ with $ \{\zg_n\}\in l^2 $.
 \item The series~(\ref{eq:laNONomegaINDIP}) is termwise differentiable.

We use these properties in  order to prove that the sequence  $ \{S_n(t)\Psi_n\} $  satisfies an equality similar to~(\ref{eq:laNONomegaINDIP}), but with one term removed.
 
\item
Hence, we can iterate the procedure, and after a finite number of steps we get
\[ 
\sum _{|n|>N} \tilde \zaa_n S_n(t)\Psi_n=0 \,.
 \]
The coefficient $ \tilde \zaa_n $ is zero if and only if the original coefficient $   \zaa_n $ is zero.

We proved that $ \{S_n(t)\Psi_n\}_{|n|>N} $ is a Riesz sequence and so for every $ n $ with $ |n|>N $ we have  $  \tilde \zaa_n =0$, hence $  \zaa_n=0 $.
\item So, the series in~(\ref{eq:laNONomegaINDIP}) is in fact a finite sum, and this   implies $ \zaa_n=0 $ since we shall prove that the sequence 
$ \{S_n(t)\Psi_n\}$ is linearly independent.
\end{itemize}

Now we proceed to realize this program. 

We state a lemma which will be repeatedly used. 

\begin{Lemma}\ZLA{Lemma:regolCOEFF}
Let 
\[ 
\Phi(x,t)=\sum_{n\in \mathbb{Z}'} \zaa_m e^{i\zben t}\Psi_n\in H^1(0,T;L^2(\Gamma)) \,.
 \]
 Then, there exists $ \{\ZDE_n\}\in l^2 $ such that
 \[ 
 \zaa_n=\frac{\ZDE_n}{\zben}\,.
  \]
\end{Lemma}
For completeness, we give a proof in Appendix~\ref{subs:AppendixproofLEMMA}.

We single out from the series~(\ref{eq:laNONomegaINDIP}) those terms which correspond to indices in $ \mathcal{J} $ (if any). Let
\[ 
F(t)=\sum _{n\in \mathcal {J}} \zaa_n S_n(t)\Psi_n\qquad \mbox{if $ \mathcal {J} \neq \emptyset $}\,,\qquad  F(t)=0\quad \mbox{otherwise}\,. 
 \]
 This sum is finite and for the indices in this sum we have
 \begin{align} 
 &S_n(t)=1+(\zaa+i) t+\intt (t-r)\biggl \{
 e^{-\zaa r}\left (N''(r)+iN'(r)\right )\biggr.\\
& \left. 
 -\zaa^2\intr N_1(r-s) S_n(s)\ZD s\right \} \ZD r\,.
  \end{align}
So, $ S_n(t) $    does not depend on $ n $ when $ n\in   \mathcal{J}   $ and it is of class $ H^3 $. Hence, using $ N\in H^3 $, $ F(t) $ is a fixed $ H^3 $ function (possibly zero).

When, in the next equalities, the index of the series is not explicitly indicated, we intend that it belongs to the set $ \mathbb{Z}' \setminus \mathcal{J}$.

Using~(\ref{eq:DEFIdiGn}) and~(\ref{Eq:equaIntegrperSN-FORMsemi-finale}) we rewrite~(\ref{eq:laNONomegaINDIP})  as
\begin{eqnarray}
&&\nonumber  -\sum \zaa_n e^{i\zben t}\Psi_n = F(t)+\zaa\sum \frac{\zaa_n}{\zben} \Psi_n \sin\zben t \\
&&\nonumber  +\intt N_1(t-s)\sum \zaa_n \left ( e^{i\zben s}+\frac{\zaa}{\zben}\sin\zben s\right )\Psi_n\ZD s\\
&&\nonumber-\intt N_1(t-r)\sum \zaa_n \mu_nS_n(r)\Psi_n\ZD r \\
&& \nonumber   
  +N_1'(0)\sum \intt \frac{\zaa_n\mu_n}{\zben}\sin\zben(t-r)S_n(r)\Psi_n\ZD r \\
 &&\ZLA{eq:primaFORMAserieNONomeginD} 
  + \intt N_1''(s)\sum \frac{\zaa_n\mu_n}{\zben}\int_0^{t-s} \sin\zben(t-s-r)S_n(r)\Psi_n\ZD r\,\ZD s\,. 
\end{eqnarray}
We prove that every term on the right hand side is of class $ H^1(0,T;L^2(\Gamma)) $.

Thanks to the fact that $ \{\Psi_n\sin\zben s\} $ and $ \{\psi_n\cos\zben s\} $ are Riesz sequences in $ L^2(G_T) $ (see Corollary~\ref{Corollary:daExpAcOS})
each series on the right hand side, a part possibly the 
two last rows, is differentiable.

The
series~(\ref{eq:SarieDaconvePASSO1}) below is the series at the second last row. If it is differentiable, then also the series at the last row is differentiable.  So, we prove differentiability of
\begin{equation}
\ZLA{eq:SarieDaconvePASSO1}
\sum \frac{\zaa_n\mu_n}{\zben}\intt \sin\zben (t-r)S_n(r)\Psi_n\ZD r\,.
\end{equation}
 Convergence of this series is clear, using~(\ref{rappreEXPdiSn}) and $ d\leq 3 $.

  Now we prove that the following series, obtained by formal termwise differentiation, converges in $ L^2(G_T) $:
 \begin{equation} \ZLA{eq:DariCGHIainAPPE}
 \sum \zaa_n\mu_n \intt \cos\zben(t-r) S_n(r)\Psi_n\ZD r\,.
  \end{equation}
  We replace $ S_n(r) $ with its 
  expression~(\ref{eq:DIseqP19}) and we get the following series:
\begin{align}
\nonumber
\sum \zaa_n\mu_n \intt \cos\zben (t-s) E_n(s)\Psi_n\ZD s\\
\nonumber-i N'(0)
\sum\frac{\zaa_n }{\zben}\intt s\cos\zben(t-s) e^{i\zben s}\Psi_n\ZD s\\
\ZLA{eq:DeriPRIkam}
+\sum \frac{\zaa_n }{\zben^2}\intt M_n(s)\cos\zben(t-s)\ZD s\,.
\end{align}
The last series converges since $ d\leq 3 $ and the first and second series converge, thanks to Corollary~\ref{Corollary:daExpAcOS}, because the convolution integrals are   combinations of $ \cos\zben t$, $ \sin\zben t $ and $ e^{i\zben t} $.

  So, we have
\[ 
\sum \zaa_ne^{i\zben t}\Psi_n\in W^{1,2}(0,T;L^2(\Gamma))\,,\qquad  \zaa_n=\frac{\ZDE_n}{\zben}\,,\qquad \{\ZDE_n\}\in l^2\,.
 \]

We replace this expression of $ \{\zaa_n\} $ in~(\ref{eq:primaFORMAserieNONomeginD}) and we equate the derivatives of both the sides. We get

 \begin{align}
\nonumber &-i\sum\ZDE_n e^{i\zben t}\Psi_n = F'(t) +\zaa\sum \frac{\ZDE_n }{\zben}\Psi_n\cos\zben t\\
\nonumber &+\intt N_1'(t-s)\sum \frac{\ZDE_n}{\zben}\left (
e^{i\zben s}+  \frac{\zaa}{\zben}\sin\zben s
\right )\Psi_n\ZD s\\
\nonumber  &-\intt N_1'(t-r)\sum \frac{\ZDE_n\mu_n}{\zben}S_n(r)\Psi_n\ZD r\\
\nonumber  &+N_1'(0)\intt \sum \frac{\ZDE_n\mu_n}{\zben}\cos\zben (t-r) S_n(r)\Psi_n\ZD r\\
\ZLA{eq:perAppend}& + \intt N_1''(s)\int_0^{t-s}\sum \frac{\ZDE_n\mu_n}{\zben} \cos\zben(t-s-r)S_n(r)\Psi_n\ZD r\,\ZD s\,.
\end{align}
Arguments similar to the previous ones show that 
  every series on the right hand side can be differentiated once more. The term in the second last line is the one that deserves a bit of attention. Its derivative is
the sum of   the two series
\begin{align*}
&N'_1(0)\sum\frac{\ZDE_n\mu_n}{\zben} S_n(t)\Psi_n\,,
\\
&
-N'_1(0)\sum\ZDE_n\mu_n\intt \sin\zben(t-r)S_n(r)\Psi_n\ZD r\,.
 \end{align*}
 The first series converges thanks to the first statement in Theorem~\ref{Lemma:RieszPERnGRANDE}. 
 
 We insert~(\ref{eq:DIseqP19})  in the second series and we    get
\[ 
\sum  \ZDE_n\mu_n \Psi_n\intt \sin\zben(t-r)\left \{
E_n(r)-irN'(0)\frac{1}{\zben} e^{i\zben r}+\frac{1}{\zben^2}M_n(r)
\right \} \ZD r\,.
 \]
 Convergence of this series is seen as in the first step of differentiation.
 
 Hence   we get
 \[ 
\ZDE_n=\frac{\tilde\zg_n}{\zben}\,,\quad  \zaa_n=\frac{\tilde\zg_n}{\zben^2}\,,\qquad \{\tilde\zg_n\}\in l^2\,.
  \]
  
  Now we iterate this process: we replace $ \ZDE_n $ with $ \tilde\gamma_n/\zben $ and we equate the derivatives. We get
    \begin{equation}\ZLA{eq:FINEzaaN} 
   \tilde \zg_n=\frac{\zg_n}{\zben}\,,\quad {\rm i.e.}\quad \zaa_m=\frac{\zg_n}{\zben^3}\,,\qquad \{\zg_n\}\in l^2\,.
    \end{equation}
Details of the computations are in Appendix~\ref{subs:AppendixproofLEMMA}.

  So, we have: 
\begin{Theorem}
If the telegraph equation is controllable in time $ T $ then there exists a sequence $ \{  \zg_n\}\in  l^2 $ such that
 \[ 
\zaa_n=\zg_n\quad{\rm if}\quad n\in\mathcal{J}\,,\qquad  \zaa_n=\frac{ \zg_n}{\zben^3}\,,\quad {\rm if}\quad n\notin\mathcal{J}  
  \]
  where $ \zaa_n $ are the coefficients in the series~(\ref{eq:laNONomegaINDIP}). 
  \end{Theorem}
  
  We recall the definition of $ S_n(t) $ in terms of $ Z_n(t) $ and we rewrite~(\ref{eq:laNONomegaINDIP}) as
\begin{equation}
\ZLA{eq:omoegDEPdiZn}
\sum _{n\in\mathbb{Z}'}\zaa_n\Psi_nZ_n(t)=\sum _{n\in\mathbb{Z}'}\frac{\zg_n}{\zben^3}\Psi_nZ_n(t)=0
\end{equation}
  The series~(\ref{eq:omoegDEPdiZn}) converges uniformly so that, computing with $ t=0 $, we get:
\begin{equation}\ZLA{eq:sommaDEGLIaan}
\sum _{n\in\mathbb{Z}'}\zaa_n\Psi_n=
\sum _{n\in\mathbb{Z}'} \frac{\zg_n}{\zben^3}\Psi_n=0\,.
\end{equation}
\emph{Here $  \zg_n/\zben^2 $ has to be replaced with $ \zg_n  $ in $ n\in \mathcal{J} $. We implicitly intend this substitutions also in the next series.}

  Using $ \zben^2\asymp\zl_n^2$,  the first statement in Theorem~\ref{Lemma:RieszPERnGRANDE}   and the form of $ K_n(t) $ and $ d\leq 3 $, we get
  \begin{Corollary}
 The series~(\ref{eq:omoegDEPdiZn}) is termwise differentiable.
 \end{Corollary}
  Hence we have:
 \[ 
 \sum\zaa_n\Psi_n\left \{
 -\zl_n^2\intt N(t-s)Z_n(s)\ZD s+K_n(t)
 \right \}=0
  \]
 
We can distribute the series on the sum and we get
 
 \begin{equation}\ZLA{eq:conVEconKn}
 \intt N(t-s) \sum _{n\in\mathbb{Z}'}\frac{\zg_n\zl_n^2}{\zben^3} Z_n(s)\Psi_n\ZD s=\sum _{n\in\mathbb{Z}'}\frac{\zg_n}{\zben^3}  K_n(t)\Psi_n\,.
  \end{equation}

 Using~(\ref{eq:sommaDEGLIaan}) and  $
 K_n(t)=N'(t)+i\zben N(t)
  $
  we get
  \[ 
  \sum \frac{\zg_n}{\zben^3}K_n(t)\Psi_n=iN(t)\sum\frac{\zg_n}{\zben^2}\Psi_n
   \]
   and so
 \[ 
  \intt N(t-s) \sum _{n\in\mathbb{Z}'}\frac{\zg_n\zl_n^2}{\zben^3} Z_n(s)\Psi_n\ZD s=iN(t)\sum _{n\in\mathbb{Z}'}\frac{\zg_n}{\zben^2} \Psi_n\,.
  \]
 Computing with $ t=0 $ and using $ N(0)=1\neq 0 $ we see that $ \{\zaa_n\} $ has a further property:
 \begin{equation}\ZLA{eq:ZNnoIndiOMEG}
 \sum _{n\in\mathbb{Z}'}\frac{\zg_n}{\zben^2} \Psi_n=\sum _{n\in\mathbb{Z}'}\zben\zaa_n \Psi_n=0 
  \end{equation}
  and so the right hand side of~(\ref{eq:conVEconKn}) vanishes.
    
Using again $ N(0)\neq 0 $ in~(\ref{eq:conVEconKn}), we get
 \begin{equation}\ZLA{eq:ZNnoIndiOMEGante}
 \sum _{n\in\mathbb{Z}'}\zaa_n\zl_n^2  Z_n(t)\Psi_n=
 \sum _{n\in\mathbb{Z}'}\frac{\zg_n\zl_n^2}{\zben^3} Z_n(t)\Psi_n=0\,.
\end{equation} 
  
 We recall equality~(\ref{eq:omoegDEPdiZn}):
$
\sum_{n\in{\mathbb Z}'} \zaa_n Z_n(t)\Psi_n=0 \,.
$
 We introduce the finite (possibly empty) set of indices
 \[
 \mathcal{O}=\left \{
 n\,:\quad \zl_n=0
 \right \}\,.
 \]
 Note that if $ n\in{\cal O} $ then $ Z_n(t)=\hat Z(t) $, the same for every $ n $. We rewrite~(\ref{eq:ZNnoIndiOMEGante}) and~(\ref{eq:omoegDEPdiZn}) as (the sum on the right side is zero if $ \mathcal{O}=\emptyset $ ):
 \begin{equation}
\ZLA{eq:inizioITERAZ}
\sum _{n\notin {\mathcal O}}\zaa_n Z_n(t)\Psi_n=
-\sum _{n\in {\mathcal O}}\zaa_n Z_n(t)\Psi_n\,, \qquad
\sum _{n\notin {\mathcal O}}
\zaa_n\zl_n^2  Z_n(t)\Psi_n=0\,.
\end{equation}
  
 
 Let $ k_1\notin{\cal O} $ be an index (of minimal absolute value) for which $ \zaa_{k_1}\neq 0 $. Combining the equalities in~(\ref{eq:inizioITERAZ}) we get
 \[ 
 \sum _{n\notin {\mathcal O}}\left (\zaa_n-\frac{\zaa_n\zl_n^2}{\zl_{k_1}^2}\right )Z_n(t)\Psi_n=-\sum _{n\in {\mathcal O}}\zaa_n Z_n(t)\Psi_n\,.
  \]
 Note that the right hand side is the same as in the first equality of~(\ref{eq:inizioITERAZ}).
 
 Let 
 $$ 
 \zaa_n^{(1)}= \left (1- \frac {\zl_n^2}{\zl_{k_1}^2 } \right ) \zaa_n 
 $$
  and note that
  \[
   \left \{\zaa_n^{(1)}\right \}\in l^2 \,;\qquad\left\{\begin{array} {l}
      \zaa_{k_1}^{(1)}=0\ {\rm if}\ \zl_k=\zl_{k_1}\\
  \mbox{if $ \zl_k\neq \zl_{k _1}$ then $ \zaa_k^{(1)} =0\ \iff\ \zaa_k=0$.}
  \end{array}\right.
   \]
 So,
 
 \begin{equation}
\ZLA{eq:APPART1}
\left\{
\begin{array}{l}
\displaystyle \sum _{n\in {\mathcal O}}\zaa_n Z_n(t)\Psi_n\in X_1=
  {\rm cl\, span}  \left \{ 
  Z_n(t)\Psi_n\,,\quad n\notin{\cal O}\,,\ \zl_n\neq \zl_{k_1}
   \right \}\\ 
 \sum _{\stackrel{n\notin {\mathcal O}}{\zl_n\neq \zl_{k_1}}}\zaa_n^{(1)}  Z_n(t)\Psi_n=
- \sum _{n\in {\mathcal O}}\zaa_nZ_n(t)\Psi_n\,.
\end{array}\right.
\end{equation}
 Thanks to $ \left \{ \zaa_n^{(1)}\right \} \in l^2 $, we can start a bootstrap argument and  repeat  this procedure: we find that $\left \{\zl_n^3 \zaa_n^{(1)}\right \} \in l^2$. We fix a second element $ k_2  $ (of minimal absolute value)
 such that $\zaa^{(1)}_{k_2}\neq 0 $ and, acting as above, we get

  \begin{equation}
\ZLA{eq:APPART2}
\left\{
\begin{array}{l}
\displaystyle \sum _{n\in {\mathcal O}}\zaa_n Z_n(t)\Psi_n\in X_2=
  {\rm cl\, span}  \left \{ 
  Z_n(t)\Psi_n\,,\quad n\notin{\cal O}\,,\ \zl_n\notin\{ \zl_{k_1}\,,\ \zl_{k_2}\}
   \right \}\\ 
 \sum _{\stackrel{n\notin {\mathcal O}}{\zl_n\notin\{ \zl_{k_1}\,,\ \zl_{k_2}\}}}\zaa_n^{(2)}  Z_n(t)\Psi_n=
 \sum _{n\in {\mathcal O}}\zaa_nZ_n(t)\Psi_n\,.
\end{array}\right.
\end{equation}
 The new sequence $ \left \{\zaa_n^{(2)} \right \}\in l^2 $ has the property that
 \[ 
 \left\{
 \begin{array}
 {l}
  \zaa_{k }^{(2)}=0\ {\rm if}\ \zl_k\in \{ \zl _{k_1}\,, \ \zl _{k_2} \}\\
   \mbox{if $ \zl_n\notin\{\zl_{k_1}\,,\ \zl_{ k_2}\} $ then $ \zaa_n^{(2)} =0\ \iff\ \zaa_n=0$.}
 \end{array}
 \right.
%
  \]
 
 Repeating this argument, we find
 \[ 
 \sum _{n\in {\mathcal O}}\zaa_n Z_n(t)\Psi_n\in X_R=
  {\rm cl\, span}  \bigl \{ 
  Z_n(t)\Psi_n\,,\quad n\notin{\cal O}\,,\ \zl_n\notin\{\zl_{ k_1}\,,\ \zl_{k_2}\,,\dots \zl_{k_R}\}
  \bigr \}
  \]
  for every $ R $, i.e. 
  
 \begin{Lemma} We have:
 \[  
  \sum _{n\in {\mathcal O}}\zaa_n Z_n(t)\Psi_n\in\bigcap_{R} X_R=\{0\}
 \]  
 and,    after at most $ 2N $ iteration of the process, we find
 \[ 
 \sum _{|n|>N}\zaa_n^{(N)}  Z_n(t)\Psi_n=0\,.
  \]
 \end{Lemma}
 If $ N $ is large enough, as specified in the first statement of Theorem~\ref{Lemma:RieszPERnGRANDE}, we see that
 \[ 
 \zaa_n^{(N)} =0 \quad \mbox{for all $ n>N $} 
  \]
  and the original equality~(\ref{eq:omoegDEPdiZn}) involves a finite sum. We rewrite it as
  \begin{equation}
\ZLA{eq:combiFINI}
\sum_{\stackrel{|n|\leq K}{n\notin{\cal O}} }\zaa_n Z_n(t)\Psi_n=0\,.
\end{equation}

This equality implies $ \zaa_n=0 $ for every $ n\notin\mathcal {O} $ since
\begin{Lemma}
The sequence $ \left \{ Z_n(t)\Psi_n(x)\right \} _{n\notin\mathcal{O}} $ is linearly independent.
\end{Lemma}
\zProof 
Lemmas~\ref{Corollary:primosulbordo} and~\ref{lemma:perTRACCIAcombi} show that $ \Psi_n(x)\neq 0 $ so that we can confine ourselves to prove that $ \{Z_n(t)\} $ is linearly independent.
The proof is similar to the proof of the corresponding result in~\cite{AvdoninPANDOLFI1,PandVelDeformTRACT}  and is omitted.\zdia

In conclusion, Equality~(\ref{eq:omoegDEPdiZn}) is in fact
\[ 
0=\sum _{n\in\mathcal O}\zaa_n Z_n(t)\Psi_n  =\tilde Z(t)
\sum _{n\in\mathcal O}\zaa_n  \Psi_n\,,\qquad \tilde Z(t)\neq 0\,.
 \]
 So,
 \[ 
 \sum _{n\in\mathcal O}\zaa_n  \Psi_n=0\,.
  \]
 
Finally we prove:
 \begin{Lemma}
 If $ n\in \mathcal{O} $ then $ \zaa_n=0 $.
 \end{Lemma}
 \zProof We introduce
 \[ 
 \Phi(x)=\sum _{n\in\mathcal{O}}\zaa_n\Phi_n(x)
  \]
  which is an eigenfunction of the operator $ A $ whose eigenvalue is $ 0 $
  \[ 
  A\Phi(x)=0\,.
   \]
 Note that if $ \zl_n=0 $ then $ \zben=i\zaa $ does not depend on $ n $ and so
 \[ 
 \Psi_n=\left\{\begin{array}{lll}
\displaystyle  \frac{\zg _a\Phi_n}{\zben}=\frac{\zg _a\Phi_n}{i\zaa}
 &{\rm if}& \zaa\neq 0
 \\
\displaystyle  \zg_a\Phi_n&{\rm if}& \zaa= 0\,.
 \end{array}\right.
  \]
So, in both the cases, we get
\[ 
A \Phi=0\,,\qquad \zg_a\Phi=0\,.
 \]
Using Lemma~\ref{lemma:perTRACCIAcombi}, we see that
 \[ 
 0=\Phi(x)=\sum \zaa_n\Phi_n(x)=0\,.
  \]
  The condition $ \zaa_n=0 $ follows, since $ \{\Phi_n\} $ is an orthonormal sequence.\zdia

\section{\ZLA{section:SHARP}Sharp control time}

It is clear that we can't control \emph{every} system of the form  ~(\ref{eq:SisteMEMsecondordine}) at a ``small'' time $ T_0 $, at which the corresponding telegraph equation is not controllable, since $ M(t)=0 $ is a possible choice of the kernel. In this section we are going to improve this obvious observation, in that non controllability of the viscoelastic system implies non controllability of the corresponding telegraph equation, regardless of the relaxation kernel $ M(t) $.
\begin{Theorem}\ZLA{teo:converse}
Let the telegraph equation~(\ref{eq:TeleCONFRO}) be non controllable at time $ T_0 $. Then, Eq.~(\ref{eq:SisteMEMsecondordine}) is not controllable at time $ T_0 $, for \emph{every} choice of the $ H^2_{{\rm loc}}(0,+\ZIN) $ kernel $ M(t) $.
\end{Theorem}
\zProof
We proceed by contradiction: let Eq.~(\ref{eq:SisteMEMsecondordine}) be controllable at time $ T $. Then, the moment problem~(\ref{eq:MomePROcomplPERvisco}) is solvable in $ l^2 $ and then the sequence $ \{Z_n(t)\Psi_n\} $ is a Riesz sequence in $ L^2(0,T;L^2(\Gamma)) $. Using the first statement of Theorem~\ref{Lemma:RieszPERnGRANDE} the converse way around, we get that there exists a number $ N $ such that the sequence
whose elements are described in Theorem~\ref{teo:RieszPERtele}, and index $ n $ such that $|n|>  N $, is Riesz in $ L^2(0,T_0;L^2(\Gamma)) $.

This implies that the moment problem~(\ref{eq:MOmePROtelegEQUA1})-(\ref{eq:MOmePROtelegEQUA2}) \emph{for the telegraph equation} is solvable for $ \{\xi_n,\eta_n\}\in L $, where $ L $ has \emph{finite codimension,}
 see\cite[p.~323]{GohbergKrein} I.e., the reachable set at time $ T_0 $ for the telegraph equation wuold have finite codimension.We use Lemma~\ref{Lemma:delleSOTTOS} in order to prove that this is not the case.

 Let $ T>T_0 $ be any time at which \emph{the telegraph equation~(\ref{eq:TeleCONFRO})  is controllable.}
 
Let us denote $ e_n $ the elements of the sequence described in  Theorem~\ref{teo:RieszPERtele}. Adding elements of $ \left ({\rm cl\, span}\, \{e_n\}\right )^\perp $, we complete the sequence $ \{e_n\} $ to a Riesz \emph{basis } of $ L^2(0,T;L^2(\Gamma)) $. We denote $ k_n $ the added elements, so that the Riesz basis of $ L^2(0,T;L^2(\Gamma)) $ is $ \{e_n\}\cup \{k_n\} $.

We consider the operator $ {\mathbb J}_0 $: $ L^2(0,T_0;L^2(\Gamma))\mapsto l^2 $ given by
\[ 
{\mathbb J}_0  f=\bigl\{\langle f,e_n\rangle_{L^2(0,T_0;L^2(\Gamma))}  \bigr\}\cup \bigl\{\langle f,k_n\rangle_{L^2(0,T_0;L^2(\Gamma))}\bigr\}\,.
 \]
Lemma~\ref{Lemma:delleSOTTOS} shows that the codimension of its image is not finite and so we have also
\[ 
{\rm dim}\, L^\perp ={\rm dim}\, \bigl\{\langle f,e_n\rangle_{L^2(0,T_0;L^2(\Gamma))}  \bigr\}^\perp =+\ZIN\,.
 \] 
 So, the index $ N $ cannot exists and the viscoelastic system cannot be controllable at a time $ T_0 $ where the telegraph equation is not controllable.\zdia

The previous negative result has a clear relation with the following fact, that the speed of propagotion of waves in a viscoelastic material is equal to the speed of propagation in the corresponding memoryless elastic material, see~\cite{FisherGURTIN,HerreraGURTIN}.

\section{Appendix: proofs}

in this appendix we prove Lemma    and the last step of differentiation.

\subsection{\ZLA{subs:AppendixproofLEMMA}Appendix~1: the proof of Lemma~\ref{Lemma:regolCOEFF}}

We first note that Theorem~\ref{Lemma:RieszPERnGRANDE} implies  $ \{\zaa_n\}\in l^2 $ and that in order to prove the formula for $ \{\zaa_n\} $ it is sufficient that we prove that it holds for $ |n| $ sufficiently large. So, we consider the new function
\begin{equation}\ZLA{eq:appe1:formaPhi}
C(x,t)=\sum_{|n|\geq N} \zaa_n e^{i\zben t}\Psi_n\in W^{1,2}(0,T;L^2(\Gamma))
 \end{equation}
where $ N $ is the number specified in Theorem~\ref{Lemma:RieszPERnGRANDE}.
It is known that $ C'(x,t)=C_t(x,t) $ is the $ L^2(0,T;L^2(\Gamma)) $ limit of the incremental quotient respect to the variable $ t $:
\[ 
C_t(x,t)=\lim _{h\to 0}\frac{C(x,t+h)-C(t,x)}{h}
=\lim _{h\to 0}
\sum _{|n|>N} \zaa_n \frac{e^{i\zben h}-1}{h}e^{i\zben t}\Psi_n\,.
 \]
 Thanks to the choice of $ N $, there exists $ m_0>0 $ such that
 \begin{align*}
 &m_0\sum _{|n|>N}\left |
 \zaa_n\zben \frac{e^{i\zben h}-1}{ \zben h}
 \right |^2\leq \left \| \frac{C(x,t+h)-C(x,t)}{h}\right \|^2_{L^2(0,T;L^2(\Gamma))}
 \\&
 \leq 2\|C'\|^2_{L^2(0,T;L^2(\Gamma))}\,.
 \end{align*}
 The last equality holds for $ h $ ``small'', $|h|<h_0$. We consider $ 0<h<h_0 $.
 
Let $ s $ be real. There exists $ s_0>0 $ such that: 
 \[ 
 \left |  \frac{e^{is}-1}{s}\right | ^2=\left ( \frac{\cos s-1}{s} \right )^2+\left ( \frac{\sin s}{s}  \right )^2>\frac{1}{2}\quad \mbox{for $ 0<s<s_0 $.}
  \]
Then we have, for every $ h\in (0,h_0)$,

\begin{align*}
\frac{1}{2}\sum _{\stackrel{|n|>N}{\ \zben<s_0/h}  }|\zaa_n\zben|^2\leq 
\sum _{|n|>N}\left |
 \zaa_n\zben  \frac{e^{i\zben h}-1}{ \zben h}
 \right |^2\leq \frac{2}{m_0}\|C'\|^2_{L^2(0,T;L^2(\Gamma))}\,.
\end{align*}
The limit for $ h\to 0^+ $ gives the result.
\subsection{Appendix~2: end of the proof of formula~(\ref{eq:FINEzaaN})}

We insert
$
\ZDE_n= \tilde\zg_n/\zben 
 $
 in~(\ref{eq:perAppend}) and we equate the derivatives of both the sides. We get  
\begin{align*}
\nonumber& \sum \tilde\zg_n e^{i\zben t}\Psi_n=
F''(t)\Psi_n-\zaa\sum\frac{\tilde\zg_n}{\zben}\Psi_n\sin\zben t\\
\nonumber& +N_1'(0)\sum
\frac{\tilde\zg_n}{\zben^2}\Psi_n\left ( e^{i\zben t}+\frac{\zaa}{\zben}\sin\zben t\right )
\\
\nonumber&
+\intt N_1''(t-s)\sum\frac{\tilde\zg_n}{\zben^2} \Psi_n
\left ( e^{i\zben s}+\frac{\zaa}{\zben}\sin\zben s\right )\ZD s\\
 \nonumber & -N_1'(0)\sum\frac{\tilde\zg_n\mu_n}{\zben^2}\Psi_n S_n(t)
 -\intt N_1''(t-r)\sum \frac{\tilde\zg_n\mu_n}{\zben^2}\Psi_n S_n(r)\ZD r
 \\ \nonumber& 
+N_1'(0)\sum\frac{\tilde\zg_n\mu_n}{\zben^2}\Psi_n S_n(t)
%
-N_1'(0)\intt \sum\frac{\tilde\zg_n\mu_n}{\zben} \sin\zben(t-r)\Psi_n S_n(r)\ZD r
\\
\nonumber &
 +\intt N_1''(t-s) \sum\frac{\tilde\zg_n\mu_n}{\zben^2}\Psi_n S_n(s)\ZD s\\
\nonumber&-\intt N_1''(s)\int_0^{t-s}\sum \frac{\tilde\zg_n\mu_n}{\zben}  \sin\zben(t-s-r)\Psi_nS_n(r)\ZD r\,\ZD s\,.
\end{align*}

The integrals which do not contain $ S_n(t) $ can be differentiated (use the substitution $ t-s=r $ to remove the variable $ t $ from $ N_1''(t) $).

We cancel similar terms with opposite sign and we consider the remaining terms which contains $ S_n(t) $, i.e. 
\begin{align*}
&N_1'(0)\intt \sum\frac{\tilde\zg_n\mu_n}{\zben} \sin\zben(t-r)\Psi_n S_n(r)\ZD r\\
&
\intt N_1''(s)\int_0^{t-s}\sum \frac{\tilde\zg_n\mu_n}{\zben}  \sin\zben(t-s-r)\Psi_nS_n(r)\ZD r\,\ZD s\,.
\end{align*}

  We compute the derivative of the first series termwise and we see that the series so obtained is $ L^2 $-convergent:
\[ 
\frac{\ZD}{\ZD t}\intt \sum\frac{\tilde\zg_n\mu_n}{\zben}\sin\zben(t-r)S_n(r)\ZD r= 
\intt \sum \tilde\zg_n\mu_n\cos\zben(t-r) S_n(r)\ZD r\,.
 \]
 Convergence of a series of this kind has already been proved, see the proof of the convergence of the series~(\ref{eq:DariCGHIainAPPE}).
 
 The convergence of this series implies also that the last integral can be differentiated.

\enddocument